\documentclass{ws-procs9x6-cpt19}
\begin{document}

\newcommand{\refeq}[1]{(\ref{#1})}
\def\etal {{\it et al.}}

\def\al{\alpha}
\def\be{\beta}
\def\ga{\gamma}
\def\de{\delta}
\def\ep{\epsilon}
\def\ve{\varepsilon}
\def\ze{\zeta}
\def\et{\eta}
\def\th{\theta}
\def\vt{\vartheta}
\def\io{\iota}
\def\vka{\varkappa}
\def\ka{\kappa}
\def\la{\lambda}
\def\vpi{\varpi}
\def\rh{\rho}
\def\vr{\varrho}
\def\si{\sigma}
\def\vs{\varsigma}
\def\ta{\tau}
\def\up{\upsilon}
\def\ph{\phi}
\def\vp{\varphi}
\def\ch{\chi}
\def\ps{\psi}
\def\om{\omega}
\def\Ga{\Gamma}
\def\De{\Delta}
\def\Th{\Theta}
\def\La{\Lambda}
\def\Si{\Sigma}
\def\Up{\Upsilon}
\def\Ph{\Phi}
\def\Ps{\Psi}
\def\Om{\Omega}
\def\cA{{\cal A}}
\def\cB{{\cal B}}
\def\cC{{\cal C}}
\def\cD{{\cal D}}
\def\cE{{\cal E}}
\def\cH{{\cal H}}
\def\cl{{\cal L}}
\def\cL{{\cal L}}
\def\cO{{\cal O}}
\def\cV{{\cal V}}
\def\cP{{\cal P}}
\def\cR{{\cal R}}
\def\cS{{\cal S}}
\def\cT{{\cal T}}

\def\pt{\phantom}

\def\half{{\textstyle{1\over 2}}}
\def\ol{\overline}
\def\prt{\partial}

\def\Re{\hbox{Re}\,}
\def\Im{\hbox{Im}\,}

\def\lsim{\mathrel{\rlap{\lower4pt\hbox{\hskip1pt$\sim$}}
    \raise1pt\hbox{$<$}}}
\def\gsim{\mathrel{\rlap{\lower4pt\hbox{\hskip1pt$\sim$}}
    \raise1pt\hbox{$>$}}}

\def\etal{{\it et al.}}

\def\vev#1{\langle {#1}\rangle}
\def\expect#1{\langle{#1}\rangle}
\def\bra#1{\langle{#1}|}
\def\ket#1{|{#1}\rangle}

\def\tr{{\rm tr}}

\newcommand{\beq}{\begin{equation}}
\newcommand{\eeq}{\end{equation}}
\newcommand{\bea}{\begin{eqnarray}}
\newcommand{\eea}{\end{eqnarray}}
\newcommand{\rf}[1]{(\ref{#1})}
\newcommand{\nn}{\nonumber\\}

\def\mn{{\mu\nu}}

\def\kad{(k^{(d)}){}}
\def\hk{(\widehat{k}){}}
\def\hu{\widehat{u}{}}
\def\uu{\overline{u}}
\def\yy{\overline{y}}
\def\K{\widetilde{k}^{(d)}{}}
\def\ha{(\widehat{k}_a){}}
\def\hc{(\widehat{k}_c){}}

\def\cd{(k_c^{\hskip-1pt(d)})^\mn{}}
\def\cf{(k_c^{(4)})}

\title{Lorentz Violation and Riemann-Finsler Geometry}

\author{Benjamin R.\ Edwards}

\address{Physics Department, Indiana University, \\
Bloomington, Indiana 47405, USA}

\begin{abstract}
The general charge-conserving
effective scalar field theory
incorporating violations of 
Lorentz symmetry is presented.
The dispersion relation 
is used to infer the effect 
of spin-independent
Lorentz violation on
point particle motion.
A large class of
associated Finsler spaces 
is derived,
and the properties of
these spaces are explored.
\end{abstract}

\bodymatter

\section{Introduction}
Connections between Riemann-Finsler spaces and theories
with Lorentz violation have recently been uncovered.\cite{AK04}
A lack of physical examples
is an obstacle on the path toward developing
a strong intuition about Finsler spaces.
In the first section,
the general effective 
quadratic scalar field theory incorporating
violation of Lorentz symmetry
will be developed.
In the next section,
a method to generate the lagrangian
describing the motion of an analogue point particle
experiencing spin-independent Lorentz violation
is derived.
The last section explores the properties of
these Finsler spaces.
These proceedings are based on results
in Ref.\ \refcite{BE18}.

\section{Field theory}
For a complex scalar field $\ph$
propagating in an
$n$-dimensional Minkowski spacetime 
with metric $\et_\mn$,
the quadratic Lagrange density
incorporating Lorentz violation is
\beq
\cl(\ph,\ph^\dagger) =
\prt^\mu\ph^\dagger\prt_\mu\ph - m^2\ph^\dagger\ph 
+ \half
\big[
\prt_\mu \ph^\dagger \hc^{\mn} \prt_\nu \ph
- i \ph^\dagger\ha^\mu \prt_\mu \ph 
+\text{ h.c.}
\big] 
\label{complexL} .
\eeq
The Lorentz violation is realized
by the CPT-odd operator $\ha^\mu$,
and the CPT-even operator $\hc^\mn$,
each of which can include
coefficients for Lorentz violation
associated with operators of
arbitrarily large mass dimension $d$.
The hermiticity of $\cl$
implies these operators 
are themselves hermitian.
In the special case of
hermitian scalar fields,
the term involving $\ha^\mu$
is proportional to a total derivative.
It follows that 
CPT symmetry is guaranteed when
$\ph = \ph^\dagger$.

Field redefinitions
can eliminate any traces
present in the 
coefficients for Lorentz violation
by absorbing them into the
terms with lower mass dimension.
We can therefore
take them to be traceless 
without loss of generality.
The commutativity of derivatives
implies that they are
totally symmetric in
all their indices.
From these considerations,
it is found that
the coefficients contain
$(2d - n +2)(d-1)! / (d - n + 2)(n-2)!$
independent components.

The dispersion relation
for this theory is found to be
\beq
p^2 - m^2
+ \hc^\mn p_\mu p_\nu
- \ha^\mu p_\mu
=0 ,
\eeq
where the operators
$\hc^\mn$ and $\ha^\mu$
are expressed in momentum space as
\bea
\hc^\mn &=& 
\sum_{d=n}^\infty
\cd^{\al_1\al_2\cdots\al_{d-n}}
p_{\al_1} p_{\al_2} \cdots p_{\al_{d-n}} , \nn
\ha^\mu &=&
\sum_{d=n-1}(k_a^{(d)})^{\mu{\al_1}\cdots{\al_{d-n+1}}}
p_{\al_1}p_{\al_2}\cdots p_{\al_{d-n+1}} ,
\eea
with the sums running
over even powers of $p$.
For brevity,
both types of coefficients will
be expressed without the $a$ or $c$
subscripts in what follows, 
and the appropriate sign difference 
will be absorbed into the $k^{(d)}$
coefficient where the CPT properties
will be determined by the mass dimension $d$.

\section{Classical kinematics}
A method has been developed
to extract point particle lagrangians
from a given field theory.\cite{RK10}
Using the three equations
\bea
R(p) = 0 , \\
\frac{\prt p_0}{\prt P_j}
= -\frac{u^j}{u^0} , \\
L = -u^\mu p_\mu ,
\eea
the idea is to identify the centroid
of a localized wave packet with 
the point particle.
The method starts with the 
dispersion relation $R(p)$.
Next,
the components of the classical velocity $u^\mu$
of the particle are related to the group velocity
of the corresponding wave packet.
The last equation is required 
by translation invariance of $L$,
along with the requirement that $L$
be one-homogeneous in the velocity,
$L(\la u) = \la L(u)$.
The first two relations can then be used
to eliminate the components of the $n$-momentum
to write $L$ only as a function of the velocity $u^\mu$.

These equations can be combined to
produce a quadratic polynomial in $L$.
For the case $d=n$,
solving this quadratic
leads to the exact lagrangian.
For the nonminimal cases $d \ge 5$,
corrections to the usual lagrangian
can be determined through an iterative procedure.
The process begins with an expansion
in $\kad$ of the roots of the quadratic.
Call this root $L=L(u^\mu , p_\mu , \kad )$.
Define $L_0 = L(u^\mu , p_\mu , 0) = -\sqrt{u^\mu\et_\mn u^\nu}$,
and then $L_j = L(u^\mu , p^{(j-1)}_\mu (u^\mu) , \kad )$
where $p^{(j)}_\mu = -\prt L_j / \prt u^\mu$.
This leads to
\bea
L^{(d)}_3 
&=&
L^{(d)}_0
\big[
1 - \half \K - \tfrac 18 (d-n+1)^2(\K)^2
\nn
&& 
+ \tfrac 18 (d-n+2)^2 ~\K_\al ~\K^\al 
- \tfrac 1{16} (d-n+1)^4(\K)^3
\nn
&&
+ \tfrac 1{16} (d-n+1)(d-n+2)^2(2d-2n+1)~\K ~\K_\al ~\K^\al
\nn
&&
- \tfrac 1{16} (d-n+1)(d-n+2)^3~\K_\al ~\K^{\al\be}~\K_\be 
\big],
\label{l3}
\eea
where the
\bea
\K
&=& 
m^{n-d}(k^{(d)})_{\al_1\al_2\cdots\al_{d-n+2}}\hat{u}^{\al_1}\hat{u}^{\al_2}\cdots \hat{u}^{\al_{d-n+2}} , \nn
\K_{\al_1\cdots\al_l}
&=&
m^{n-d}(k^{(d)})_{\al_1\cdots\al_l\al_{l+1}\cdots\al_{d-n+2}}\hat{u}^{\al_{l+1}} \cdots \hat{u}^{\al_{d-n+2}} ,
\eea
contain the directional dependence $\hat{u}^\mu = u^\mu / u$,
$u = \sqrt{u^\mu\et_\mn u^\nu}$.
This Lagrangian matches
the first order correction
found by Reis and Schreck for the nonminimal
fermion sector using an ansatz-based technique.\cite{RS18}

\section{Finsler geometry}
The Finsler structure (or Finsler norm)
plays a central role in the study of Finsler spaces.
Classical lagrangians satisfy many
of the requirements in the definition of Finsler structures.
Though there are important differences preventing the
lagrangians derived above from fulfilling 
the definition of a Finsler structure,
a method exists to generate associated Finsler structures
from a given lagrangian.\cite{AK11}

For the lagrangians developed from
the scalar field theories discussed above,
the prescription to generate a Finsler structure
in this case is given by
$p_x(u) \rightarrow (-i)^n p_x(y)$,
$k^{(d)x} \rightarrow i^nk^{(d)x}$,
$L \rightarrow -F = -y\cdot p$, 
$u^x \rightarrow (i)^n y^x$.

As a demonstration of the kinds
of geometrical quantities one can
calculate in these spaces,
we use the Finsler space
associated with the first order
limit of the lagrangian given in Eq.\ (\ref{l3}).
The Finsler structure associated with this lagrangian is
\beq
F^{(d)}
= y - \half y \K .
\eeq

The fundamental tensor of a Finsler space
determines the metric and therefore
also the geodesics.
The definition of this tensor is
$g_{jk} = \half \prt^2 F^2 / \prt y^j \prt y^k$.
For the limit under consideration,
the fundamental tensor is given by
\bea
g^{(d)}_{jk}
&=&
r_{jk}(1+\half(d-n)\K) -\half(d-n+2)(d-n+1)\K_{jk} \nn
&+& \half(d-n)(d-n+2)[\K_j\hat{y}_k+\K_k\hat{y}_j-\K\hat{y}_j\hat{y}_k] .
\eea
Inspection shows $g_{jk}$ reduces to 
a purely riemannian one
for the cases $d=n$ and $d=n-2$.
This is consistent with the
fact that the $d=n$
coefficient can be absorbed 
into the metric at the level of the field theory,
while a $d=n-2$
coefficient would correspond to
a rescaling of the mass term.

The situation is not as straightforward
for other values of mass dimension.
It has been demonstrated that
the nonvanishing of the Cartan torsion
implies non-riemmannian nature of the underlying space.\cite{D53}
Calculation of this tensor shows
it vanishes for $d=n$ and $d=n-2$,
and also in the case of $n=1$,
which represents a Riemann curve,
but is nonvanishing in other cases.
Calculation of the Matsumoto torsion\cite{M72}
shows only $d=n-1$ reduces to a Randers metric.

The geodesics are obtained from the
geodesic equation $F\tfrac{d}{d\la}(y^j / F) + G^j =0$.
A calculation shows the geodesic spray coefficients $G^j$ are
\bea
\frac{1}{y^2}G^j
&=& \half\widetilde{D}^j\K
+ \half(d-n)\hat{y}^j\widetilde{D}_\bullet\K \nn
&-&
\half(d-n+2)r^{jl}\widetilde{D}_\bullet\K{}_l
+ \widetilde{\ga}^j{}_{\bullet\bullet} ,
\eea
where a $\bullet$
subscript denotes contraction of $\hat{y}^j$
with a lower $j$ index,
with all contractions exterior to any derivatives.

It is clear from this expression
that if the background field
is covariantly constant with respect
to the riemannian metric,
$\widetilde{D}_j \K_l=0$,
then the geodesics are unaffected.
This situation was conjectured 
to hold in general and is confirmed here,
but remains unproved.

\end{document}